\documentclass{ws-ijmpa}

\begin{document}

\markboth{Pawe{\l} Klaja and Pawe{\l} Moskal}
{Two-proton correlation function for the $pp \to pp + \eta$ and $pp \to pp + pions$ reactions}

%
\catchline{}{}{}{}{}
%

\title{TWO-PROTON CORRELATION FUNCTION FOR THE $pp \to pp + \eta$ AND $pp \to pp + pions$ REACTIONS}

\author{PAWE{\L} KLAJA$^{\star}$ and PAWE{\L} MOSKAL$^{\star,\bullet}$  for the COSY-11 collaboration}

\address{
$^{\star}$Institut f{\"u}r Kernphysik, Forschungszentrum J\"{u}lich, 
52425 J\"ulich, Germany\\
$^{\bullet}$Institute of Physics, Jagiellonian University,
30-059 Cracow, Poland\\}

\maketitle


\begin{abstract}
For the very first time, the correlation femtoscopy method is applied
to a kinematically complete measurement of meson production in the collisions of hadrons.
The shape of the two-proton correlation function
derived 
for the $pp\to pp\eta$ reaction differs from that for the $pp\to pp(pions)$ and both
do not show a peak structure opposite to
results determined for inclusive measurements
of heavy ion collisions.

\end{abstract}

\ccode{PACS numbers: 13.60.Hb, 13.60.Le, 13.75.-n, 25.40.Ve}

In this contribution we briefly describe results presented in details in reference~\cite{jphysg_corr}.

Momentum correlations of particles at small relative velocities are
widely used to study the spatio-temporal characteristics of
 the production processes
in relativistic heavy ion collisions\cite{lisa}.
This technique, called {\em correlation femtoscopy}\cite{led},
originates from photon intensity interferometry
initiated by Hanbury Brown and Twiss\cite{hbt}.
Implemented to nuclear physics\cite{led,koonin,kopyl}
it permits to
determine the duration of the emission process and the size
 of the source from which the particles are emitted\cite{led}.
In the case of the $pp\to pp\eta$ reaction the knowledge of this size
 might be essential to answer the intriguing question whether
 the three-body $pp\eta$ system is capable of supporting
 an unstable Borromean bound state postulated by Wycech\cite{wycech_acta}. 
 
  In contrast to heavy ion collisions\cite{chaj}, in the case of single meson production,
 the kinematics of all ejectiles may be entirely determined and hence
 a kinematically complete measurement of meson production
 in the collisions of hadrons
 gives access to complementary information
 which could shed light on the interpretation of the two-proton
 correlations observed in heavy ion reactions.
 It is also important to underline that the
 correlation of protons was never exploited till now in near threshold
 meson productions, and as an observable
 different from the distributions of cross sections,
 it may deepen our understanding
 of the dynamics of meson production.
 

Here we report on a $\eta$ meson and multi-pion production experiment
 in which the mesons were created
 in collisions of protons at a beam momentum of 2.0259~GeV/c\cite{prc69}.
Momentum vectors of outgoing protons from the $pp\to ppX$ reaction
were measured by means of the COSY-11 detector\cite{brauksiepe}.

\begin{figure}[h]
\parbox{0.38\textwidth}{\psfig{file=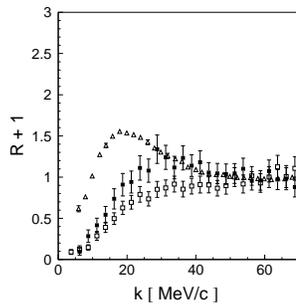,width=4.7cm}}
\parbox{0.6\textwidth}{\caption{The two-proton acceptance corrected
  correlation functions normalized to
  the corresponding simulated correlation function for a point-like source.
  Results
  for the $pp \to pp\eta$ (full squares) and $pp\to pp+pions$ (open squares)
  are compared to the two-proton correlation function determined
  from heavy ion collisions (triangles). 
  The double ratio is a well established measure of correlations
  used e.g. by the ALEPH, OPAL and DELPHI collaborations\protect\cite{aleph1,delphi,opal}.
  Variable $k$ denotes  the proton momentum in the proton-proton ceneter of mass frame.
  \label{double}}}
\end{figure}

The shape of the obtained proton-proton
correlation function reflects not only the
 space-time characteristics of the interaction volume
 but it may also be strongly modified
by the conservation of energy and momentum and by the final state interaction among the ejectiles.
In order to extract from the experimental data the shape
of the correlation function free from these effects we constructed a double ratio
dividing the experimental functions by the  
  corresponding simulated correlation function for a point-like source\cite{jphysg_corr}.
  The determined double ratios are presented in Figure~\ref{double}.
\vspace{-0.7cm}
\vspace{-0.6cm}
A significant discrepancy between the two-proton
correlation functions determined from inclusive heavy ion reactions\cite{xx}
and from exclusive proton-proton measurements\cite{jphysg_corr}
is clearly visible. The data from the kinematically
exclusive measurement do not reveal a peak structure at 20~MeV/c.

At present it is not possible to draw a solid quantitative conclusion
about the size of the system since  e.g. in the case of the $pp\to pp\eta$ reaction
it would require to solve a three-body problem
where $pp$ and $p\eta$\cite{wycech} interactions 
contribute significantly to the proton-proton correlation.
However, based on semi-quantitative predictions\cite{deloff}
one can estimate that the system must be unexpectedly large
with a radius in the order of 4~fm. This makes the result interesting in context
of the predicted quasi-bound $\eta NN$ state\cite{ueda} and in view of the
hypothesis\cite{wycech_acta}
that 
the proton-proton pair may be emitted
from a large Borromean like object whose radius is about 4~fm.



\end{document}